\documentclass[sigconf,screen]{acmart}
\AtBeginDocument{%
  }

\setcopyright{cc}
\setcctype{by}
\acmDOI{10.1145/3832783.3834340}
\acmYear{2026}
\copyrightyear{2026}
\acmISBN{979-8-4007-2882-2/2026/10}
\acmConference[ASE '26]{Proceedings of the 41st IEEE/ACM International Conference on Automated Software Engineering}{October 12--16, 2026}{Munich, Germany}
\acmBooktitle{Proceedings of the 41st IEEE/ACM International Conference on Automated Software Engineering (ASE '26), October 12--16, 2026, Munich, Germany}
\acmSubmissionID{ase26main-p135-p}
\received{2026-03-26}
\received[accepted]{2026-06-18}

\usepackage{xcolor}
\usepackage{makecell}
\usepackage{microtype}

\usepackage{graphicx} 

\usepackage{booktabs}

\usepackage{array}     

\usepackage{multirow}    
\usepackage[most]{tcolorbox}

\begin{document}

\newtcolorbox[auto counter]{summary}[1][]{
        title={\bfseries Summary},enhanced,
	coltitle=black,
	top=0.17in,
	attach boxed title to top left=
	{xshift=1.5em,yshift=-\tcboxedtitleheight/2},
        boxed title style={size=small,colback=lightgray},#1}

\newcommand{\rqbox}[1]{
\begin{tcolorbox}[tile, size=fbox, boxsep=2mm, boxrule=0pt, top=0pt, bottom=0pt,
borderline west={1mm}{0pt}{blue!50!white}, colback=blue!5!white]
#1
\end{tcolorbox}
}

\title{Source-Free Detection and Impact Analysis of Compiler Optimization Problems in Mobile Applications}

\author{Han Hu}
\orcid{0000-0003-4735-2241}
\affiliation{%
  \institution{Beihang University}
  \country{China}
}
\email{agmaiofhuhan@gmail.com}

\author{Xiaoheng Xie}
\orcid{0009-0006-9700-0542}
\affiliation{%
  \institution{Huawei Hong Kong Research Center}
  \country{Hong Kong}
}
\email{xiexiemyself@gmail.com}

\author{Bo Sun}
\orcid{0009-0004-1960-3503}
\affiliation{%
  \institution{Huawei}
  \city{Wuhan}
  \country{China}
}
\email{361606685@qq.com}

\author{Jian Gu}
\orcid{0000-0002-9105-5423}
\affiliation{%
  \institution{Monash University}
  \city{Melbourne}
  \country{Australia}
}
\email{jian.gu@monash.edu}

\author{Gang Fan}
\orcid{0000-0002-8633-6036}
\affiliation{%
  \institution{Huawei Hong Kong Research Center}
  \country{Hong Kong}
}
\email{fan.gang.cn@gmail.com}

\author{Li Li}
\authornote{Corresponding author.}
\orcid{0000-0003-2990-1614}
\affiliation{%
  \institution{Beihang University}
  \country{China}
}
\email{lilicoding@ieee.org}

\renewcommand{\shortauthors}{Han Hu, Xiaoheng Xie, Bo Sun, Jian Gu, Gang Fan, and Li Li}

\begin{abstract}
  Mobile apps frequently suffer from frame drops, overheating, and excessive power consumption. While developers optimize algorithms and debug code, a critical bottleneck often goes unnoticed: native libraries compiled with low optimization levels (O0/O1 instead of O2/O3). Because these libraries execute without functional errors, the resulting performance degradation remains hidden in production apps.
  
  We present \textsc{OptDetect}, a source-free framework that detects compiler optimization problems directly from app binaries. \textsc{OptDetect} handles mixed optimization levels through binary disassembly, chunk-level classification, and weighted score aggregation, achieving 93.0\% accuracy on controlled datasets and 81.9\% on real-world datasets. Applying \textsc{OptDetect} to 21,972 native libraries from 830 top-ranked Google Play apps, we find that 30.5\% of libraries use low optimization levels, affecting 91.7\% of apps.

  Through case studies on 12 production apps, fixing detected issues reduces CPU instructions by 10-63\% (median: 20.5\%) for commercial apps and 15-58\% (median: 32\%) for open-source apps. Performance complaints decrease in 5 of 6 commercial apps, and ratings increase in 5 of 6. Further investigation reveals that widely-used third-party libraries are themselves distributed at low optimization levels, with 49.7\% of 1,073 libraries in a major repository exhibiting this problem. These findings show that compiler optimization problems are common, source-free detectable, and practically consequential in mobile app ecosystems.
\end{abstract}



\maketitle

\section{Introduction}
Mobile applications (apps) have become ubiquitous computing platforms, with over 7.2 billion devices worldwide~\cite{statista_smartphone_users_2024} executing performance-critical native code. Performance degradation, including frame drops, excessive energy use, and thermal throttling, directly affects user experience: nearly 90\% of users abandon apps due to poor performance~\cite{appdynamics_app_attention_2014}, and over 50\% uninstall apps within 30 days~\cite{clevertap_uninstalls_2019}. Our analysis reveals a neglected factor behind such issues: inappropriate compiler optimization levels in native libraries (.so files).

We first encountered this issue while debugging a popular mobile game (Game A, with over one million active users). Game A contains 106 native libraries, and users reported that its HarmonyOS~\cite{harmonyos_official} version lagged behind the Android version. Together with the development team, we confirmed that both versions shared the same app logic and native code.

\begin{figure}[t]
\centering
\includegraphics[width=0.9\linewidth]{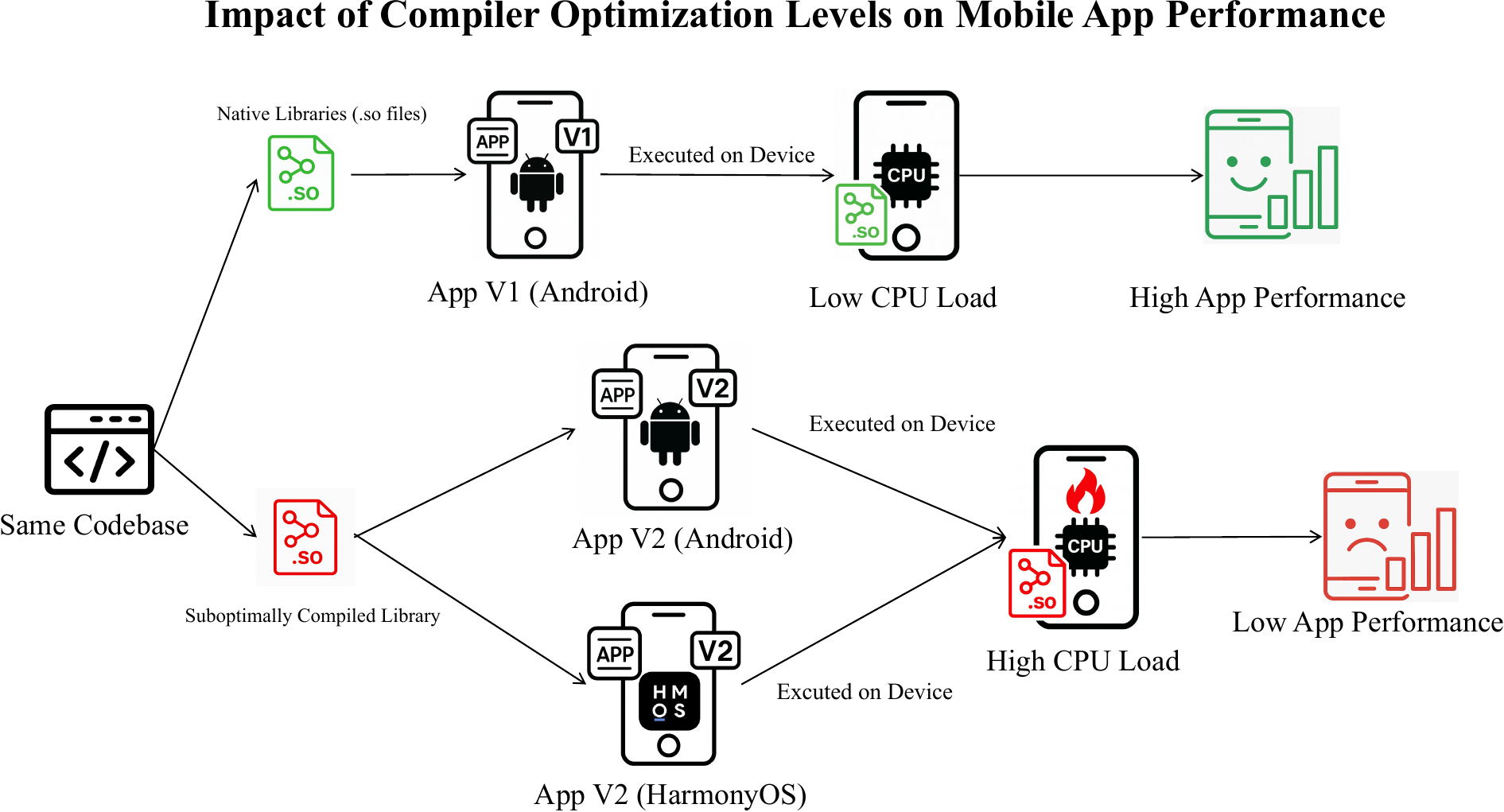}
\caption{Impact of compiler optimization levels on mobile app performance. From the same codebase, App V1 (Android) packages high-optimization libraries (O2/O3), resulting in low CPU load and high performance. App V2 (Android) and App V2 (HarmonyOS) package low-optimization libraries (O0/O1), leading to high CPU load and degraded user experience.}
\Description{A workflow diagram showing the same codebase producing Android and HarmonyOS apps with different native-library optimization levels, leading to different CPU load and app performance.}
\label{fig:optimization-issue}
\end{figure}

Runtime profiling showed that the HarmonyOS version executed approximately 60\% more CPU instructions than its Android counterpart in certain functionalities. The primary difference came from compiler settings: some HarmonyOS libraries used lower optimization levels (\texttt{O0}/\texttt{O1}), whereas the Android version used higher levels (\texttt{O2}/\texttt{O3}). Figure~\ref{fig:optimization-issue} illustrates how such differences can arise across app versions and platform builds, then propagate to runtime performance.

This case reflects a broader challenge: native libraries are often developed and compiled independently by multiple teams or third-party vendors. Fragmented build pipelines, legacy code, misconfigurations, and poor communication can therefore lead to mismatched optimization levels.

Identifying these problems in deployed apps presents challenges that existing performance tools cannot address~\cite{tang2021systematical,yang2015gator}. Profilers expose runtime symptoms, but not the optimization levels of native libraries. Prior binary classifiers such as BinEye~\cite{yang2019bineye} do not target mixed-optimization mobile libraries distributed through app stores. Since app stores expose only compiled binaries, detection must work directly from machine code and handle .so files that mix code compiled under different settings.

To address these challenges, we present \textsc{OptDetect}, an automated source-free framework that extracts native libraries from app packages and derives library-level optimization assessments through disassembly, chunk-level classification, and weighted aggregation.

We apply \textsc{OptDetect} at scale to answer three research questions: (1) How accurate and efficient is the framework? (2) How widespread is low compiler optimization in modern mobile apps? (3) What are the impacts of fixing detected problems in production apps?

\textsc{OptDetect} achieves 93.0\% accuracy on controlled datasets and 81.9\% on real-world datasets. We analyze 21,972 native libraries from 830 top-ranked Google Play apps and find that 30.5\% of libraries use low optimization levels, affecting 91.7\% of apps. Fixing detected issues in 12 production apps reduces CPU instructions by 10-63\% for commercial and 15-58\% for open-source apps. Root cause analysis identifies precompiled third-party libraries as a major source.

In summary, the main contributions of this paper are:
\begin{itemize}
    \item A source-free detection framework, \textsc{OptDetect}, that identifies compiler optimization problems directly from app binaries through a pipeline of chunk-level classification and weighted score aggregation, representing the first such framework designed for large-scale mobile app analysis without source code.
    \item The first large-scale empirical study of compiler optimization levels in mobile apps, covering 21,972 native libraries from 830 top-ranked Google Play apps.
    \item The first multi-dimensional impact analysis of compiler optimization fixes, spanning technical performance (CPU instructions, binary size, power), user-perceived quality (app store ratings and feedback), and ecosystem-level root cause investigation.
\end{itemize}

\section{Background}

Mobile apps rely on native libraries (.so files) for performance-critical functionality. These libraries are compiled by GCC~\cite{gcc} or Clang~\cite{clang} at optimization levels that control how aggressively the compiler transforms source code~\cite{muchnick1997advanced}. \texttt{O0} disables optimization for debugging, \texttt{O1} applies basic optimizations, \texttt{O2} enables more aggressive transformations such as loop and register optimizations, \texttt{O3} adds techniques such as aggressive inlining and vectorization, and \texttt{Os} optimizes for size. We treat \texttt{O0}/\texttt{O1} as low optimization levels and \texttt{O2}/\texttt{O3} as high optimization levels~\cite{muchnick1997advanced}.

Optimization level directly affects app performance: code compiled at \texttt{O0} can require substantially more CPU instructions than the same code at \texttt{O3}~\cite{muchnick1997advanced}. Yet standard profiling tools (Systrace~\cite{systrace}, Perfetto~\cite{perfetto}, DevEco Profiler~\cite{deveco}) identify CPU hotspots and memory leaks, but not whether poor performance stems from inappropriate compilation settings. This gap is especially problematic for precompiled third-party libraries, where developers lack source code or build configurations.

\section{RQ1: Detection Framework and Validation}

\subsection{Framework Design}
\label{sec:framework-design}

\begin{figure*}[t]
\centering
\includegraphics[width=0.82\linewidth]{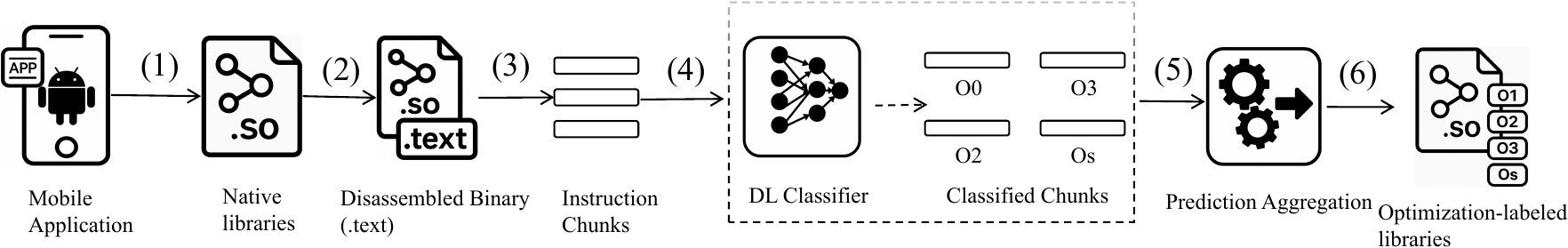}
\caption{Overview of the \textsc{OptDetect} detection framework. The six-stage pipeline consists of native library extraction, binary disassembly, instruction chunking and feature extraction, deep learning-based classification, prediction aggregation, and optimization level assignment.}
\Description{A six-stage pipeline diagram for OptDetect, from extracting native libraries and disassembling binaries to chunk-level classification, aggregation, and optimization-level assignment.}
\label{fig:approach}
\end{figure*}

Since native libraries typically contain code compiled with mixed optimization levels, accurate library-level assessment requires decomposing the analysis into (1) fine-grained chunk-level classification and (2) weighted aggregation. We realize this in \textsc{OptDetect} through a six-stage pipeline shown in Figure~\ref{fig:approach}.

\textbf{Native Library Extraction.} We unpack application packages and extract embedded native libraries (.so files) for analysis. We apply filtering criteria to exclude unsuitable binaries such as stubs, encrypted content, and heavily obfuscated binaries.

\textbf{Binary Disassembly.} We extract executable code segments from native libraries for analysis. The disassembly process focuses on the .text section, which contains the main executable code. For libraries with malformed headers, we employ full binary scanning to recover instruction sequences.

\textbf{Instruction Chunking and Feature Extraction.} To address the challenge of mixed optimization levels within libraries, we partition the .text section into fixed-size byte windows of $W$ bytes with a configurable stride $S$. Each chunk $C_i$ is a contiguous byte sequence:
\[
C_i = \{b_{i \cdot S},\; b_{i \cdot S + 1},\; \ldots,\; b_{i \cdot S + W - 1}\}
\]
where $b_j$ is the $j$-th byte in the .text section, yielding $m = \lfloor (N - W) / S \rfloor + 1$ chunks for a section of $N$ bytes. On fixed-width Instruction Set Architectures (ISAs) such as AArch64 (4-byte instructions), each $W$-byte window maps directly to $W/4$ instructions, preserving instruction-level semantics. For each chunk $C_i$, the classifier learns a latent feature representation $\mathbf{f}_i \in \mathbb{R}^d$ that captures four categories of optimization characteristics:
\[
\mathbf{f}_i = [f_{\text{opcode}}, f_{\text{register}}, f_{\text{control}}, f_{\text{structure}}]
\]
where $f_{\text{opcode}}$ encodes instruction distribution patterns, $f_{\text{register}}$ captures register allocation strategies, $f_{\text{control}}$ reflects control flow structures, and $f_{\text{structure}}$ represents code layout characteristics such as alignment and padding. These features are not hand-crafted but implicitly learned by the neural network from raw byte sequences. The specific values of $W$ and $S$ are reported in Section~\ref{sec:implementation}.

\textbf{Deep Learning Classification.} We employ a deep learning classifier to predict optimization levels at the chunk level. For each chunk $C_i$ with learned feature representation $\mathbf{f}_i$, the classifier produces a probability distribution over five optimization levels:
\[
P(y_i | \mathbf{f}_i) = \text{softmax}(h(\mathbf{f}_i))
\]
where $y_i \in \{O0, O1, O2, O3, Os\}$ represents the predicted optimization level for chunk $C_i$, and $h(\cdot)$ is the neural network function that maps learned representations to logits.

\textbf{Prediction Aggregation.} We aggregate chunk-level predictions to compute a unified library-level optimization score:
\[
\text{Score}_{\text{lib}} = \frac{\sum_{j=0}^{4} w_j \cdot \text{Chunks}_j}{\text{Total Chunks}}
\]
where $w_j$ represents the weight assigned to optimization level $j$ (reflecting its relative runtime performance, with higher optimization levels receiving higher weights), and $\text{Chunks}_j$ denotes the number of code chunks classified as level $j$. The specific weight values are reported in Section~\ref{sec:implementation}. We also compute a confidence measure based on prediction consistency:
\[
\text{Confidence} = 1 - \frac{\text{Entropy}(\mathbf{p})}{\log_2(5)}
\]
where $\mathbf{p}$ is the normalized distribution of chunk predictions across the five optimization levels.

\textbf{Optimization Level Assignment.} We assign discrete optimization levels to libraries using a hybrid approach. For libraries where a single optimization level accounts for more than $\tau_d$ of all chunks, we directly classify the library as that dominant level (e.g., O0-dominant or O3-dominant). For mixed libraries where no single level dominates, the continuous score $\text{Score}_{\text{lib}} \in [0, 1]$ captures their position on the optimization spectrum. We partition this range into four \textit{transition zones} using empirically determined thresholds, reflecting the boundaries between adjacent optimization levels (e.g., O0/O1 mixed, O1/O2 mixed). These labels reflect a library's \textit{current optimization composition on a continuous scale}, not a binary high/low judgment. The specific threshold values and their geometric justification are reported in Section~\ref{sec:implementation}.

\subsection{Validation Setup}

We train and validate our framework using two publicly available datasets. The Optimization-Detector dataset~\cite{pizzolotto2021identifying} contains approximately 17,000 .so files compiled with five optimization levels (O0, O1, O2, O3, Os) across seven architectures using GCC and Clang. The Assemblage dataset~\cite{liu2024assemblage} contains real-world libraries compiled from GitHub repositories with known compiler flags, and each binary is distributed with its ground-truth optimization level label (O0, O1, O2, O3). Because Assemblage sources code from diverse real projects with varying coding styles, build systems, and complexity, it serves as a more challenging validation benchmark than synthetically compiled datasets. To prevent data leakage, we perform the 80/20 train/test split at the project level: all .so files derived from the same source project are assigned exclusively to either the training or test set, ensuring that the model cannot memorize project-specific code patterns.

The validation data spans 7 architectures (x86\_64, AArch64, RISC-V, PowerPC, SPARC, MIPS, ARM32). In the Optimization-Detector dataset, x86\_64 and AArch64 constitute the two largest subsets (4,071 and 3,399 samples respectively), consistent with their prevalence in mobile platforms. We use 80\% for model training and internal validation, and 20\% (approximately 3,000 samples) for final testing. Within the training portion, we reserve a project-disjoint validation split for early stopping. The final test split is not used for tuning. Separately, we collect an independent calibration set of approximately 650 labeled .so libraries from mobile apps outside the RQ1 datasets. Labels are established from known build flags, developer confirmation, or source/build-script inspection. These calibration libraries are excluded from the final RQ2/RQ3 prevalence analyses and used only for aggregation-parameter selection and robustness checks.
\subsection{Implementation}
\label{sec:implementation}

We implement \textsc{OptDetect} using Python with LIEF~\cite{lief} for binary parsing and .text section extraction, Capstone~\cite{capstone} for disassembly, and PyTorch~\cite{paszke2019pytorch} for the deep learning classifier. The classifier uses a bidirectional LSTM architecture that takes raw byte sequences as input and predicts chunk-level optimization levels. For the experiments reported in this paper, we set $W = 2048$ bytes and $S = 2048$ bytes (non-overlapping windows). On AArch64, each 2048-byte window corresponds to 512 instructions, providing sufficient context for the model to capture local optimization patterns. Libraries with fewer than 5 valid chunks are excluded from analysis.

We train a single unified model across all seven ISAs and evaluate cross-ISA generalization empirically. Our assumption is not that instruction encodings are identical across ISAs, but that compiler optimizations leave recurring statistical patterns in register use, instruction selection, and code layout. In the combined training set, AArch64 and ARM32 are the two largest subsets. We train using cross-entropy loss with Adam optimizer (learning rate = 0.001, batch size = 32) for 100 epochs with early stopping. We use standard classification metrics: accuracy, precision, recall, F1-score, and false positive rate.

The aggregation weights are $w_{\text{O0}} = 0.0$, $w_{\text{O1}} = 0.25$, $w_{\text{O2}} = 0.75$, $w_{\text{O3}} = 1.0$, and $w_{\text{Os}} = 0.5$. Os is assigned 0.5 because it optimizes for binary size rather than speed, and empirical validation shows its runtime performance falls between O1 and O2. All weight values were determined through iterative empirical analysis on the independent calibration set. For optimization level assignment, the dominance threshold is 90\%, and the transition-zone boundaries are 0, 0.15, 0.35, 0.65, and 1.0, corresponding to \textit{Near-O0 mixed} $[0, 0.15)$, \textit{O0/O1 mixed} $[0.15, 0.35)$, \textit{O1/O2 mixed} $[0.35, 0.65)$, and \textit{O2/O3 mixed} $[0.65, 1.0]$. These thresholds are calibrated to maximize separation between low-optimization (O0/O1) and high-optimization (O2/O3) libraries. The prevalence thresholds used in RQ2 and RQ3 are derived from these transition zones.

\subsection{Validation Results}

We evaluate our framework on both the Optimization-Detector dataset (artificially compiled data) and the Assemblage dataset (real-world data). Table~\ref{tab:rq1-results} presents the comparative results.

\begin{table}[h]
\centering
\setlength{\tabcolsep}{4pt}
\renewcommand{\arraystretch}{0.9}
\caption{Performance evaluation of optimization level classification. \textbf{Assemblage (Binary: Low vs High)} represents binary classification accuracy for distinguishing Low-Optimization Libraries (O0/O1) from High-Optimization Libraries (O2/O3) on the Assemblage dataset.}
\label{tab:rq1-results}
\small
\begin{tabular}{lccccc}
\toprule
\textbf{\footnotesize Dataset} & \textbf{\footnotesize Level} & \textbf{\footnotesize Prec.} & \textbf{\footnotesize Rec.} & \textbf{\footnotesize Acc.} & \textbf{\footnotesize F1} \\
    \midrule
    \multirow{5}{*}{Opt-Detector} & O0 & 95.8\% & 94.2\% & 95.1\% & 95.0\% \\
    & O1 & 89.2\% & 87.6\% & 88.8\% & 88.4\% \\
    & O2 & 93.1\% & 92.8\% & 92.9\% & 93.0\% \\
    & O3 & 94.6\% & 93.8\% & 94.3\% & 94.2\% \\
    & Os & 92.5\% & 91.7\% & 92.4\% & 92.1\% \\
    \midrule
    \multirow{4}{*}{Assemblage} & O0 & 85.3\% & 83.7\% & 84.9\% & 84.5\% \\
    & O1 & 76.8\% & 74.4\% & 76.1\% & 75.6\% \\
    & O2 & 82.1\% & 81.5\% & 81.7\% & 81.8\% \\
    & O3 & 83.9\% & 82.6\% & 83.5\% & 83.2\% \\
    \midrule
    \multicolumn{2}{l}{\textbf{Opt-Detector (Macro)}} & \textbf{93.0\%} & \textbf{92.0\%} & \textbf{92.7\%} & \textbf{92.5\%} \\
    \multicolumn{2}{l}{\textbf{Opt-Detector (Weighted)}} & \textbf{93.2\%} & \textbf{92.9\%} & \textbf{93.0\%} & \textbf{93.1\%} \\
    \multicolumn{2}{l}{\textbf{Assemblage (Macro)}} & \textbf{82.0\%} & \textbf{80.6\%} & \textbf{81.6\%} & \textbf{81.3\%} \\
    \multicolumn{2}{l}{\textbf{Assemblage (Weighted)}} & \textbf{82.5\%} & \textbf{81.1\%} & \textbf{81.9\%} & \textbf{81.8\%} \\
    \midrule
    \multicolumn{2}{l}{\textbf{Assemblage (Binary: Low vs High)}} & \textbf{91.0\%} & \textbf{90.5\%} & \textbf{90.8\%} & \textbf{90.7\%} \\
    \bottomrule
    \end{tabular}%
    \end{table}

Table~\ref{tab:rq1-results} shows that our framework achieves 93.0\% weighted accuracy and 93.1\% weighted F1 on the Optimization-Detector test set and 81.9\% weighted accuracy and 81.8\% weighted F1 on the real-world Assemblage dataset. At the per-level granularity, O0 achieves the highest accuracy on both datasets (95.1\% on Opt-Detector, 84.9\% on Assemblage) due to its distinct unoptimized instruction patterns. O3 performs second best (94.3\% / 83.5\%). O1 shows the lowest performance (88.8\% / 76.1\%) due to overlap with adjacent levels, particularly on Assemblage where mixed optimization makes level boundaries ambiguous. O2 (92.9\% / 81.7\%) and Os (92.4\%) perform consistently across both datasets.

\textbf{Generalization analysis.} The approximately 11\% accuracy drop from controlled (93.0\%) to real-world (81.9\%) reflects the difference between libraries with single optimization levels and those containing mixed optimization chunks. Intermediate levels (O1, O2) remain the most challenging due to similar instruction patterns, while extreme levels (O0, O3) are more distinguishable.

\textbf{Baseline scope.} No existing method directly matches our target setting of source-free analysis for mixed-optimization mobile libraries. BinEye~\cite{yang2019bineye} is the closest prior binary-level classifier, but its evaluation setting differs from ours and it does not target mixed-optimization libraries at app scale. We therefore construct a lightweight rule-based baseline as a sanity check for whether simple binary signatures are sufficient, and discuss BinEye qualitatively in Related Work.

\begin{sloppypar}
\textbf{Comparison with rule-based baseline.} The baseline extracts five whole-binary signals: instruction density, opcode distribution (O2/O3 emits more SIMD (Single Instruction, Multiple Data) and pipeline-filling instructions), frame-pointer usage, debug-section presence, and text-to-symbol-count ratio. Each signal casts a low/high-optimization vote using fixed thresholds selected on the calibration set, and the final label is produced by majority vote. This baseline represents the family of keyword/rule/statistical checks, but it is intentionally whole-binary: it cannot localize differently optimized regions inside the same library.
\end{sloppypar}

We evaluate both methods on two regimes using the same test splits. On single-optimization-level libraries from the Opt-Detector test set (20\%), the rule-based baseline achieves 79.8\% binary classification accuracy, while \textsc{OptDetect} achieves 93.0\% (5-class), confirming that both approaches can detect optimization levels when libraries are compiled uniformly. On 100 mixed-optimization libraries randomly sampled from the Assemblage test set, whose ground truth was established through developer confirmation, build scripts, and source code inspection, the rule-based baseline drops to 68.7\% (a decline of 11.1 percentage points) because global statistics average across locally-varying chunks and cannot resolve mixed optimization patterns. \textsc{OptDetect} maintains 86.0\% accuracy on the same set, slightly above its 81.9\% on the full Assemblage test set. This comparison validates the core design choice. Chunk-level decomposition is essential for the mixed-optimization scenario that dominates real-world mobile libraries.

\textbf{Binary classification performance.} While five-class classification achieves 81.9\% on the Assemblage dataset, the binary task of distinguishing Low-Optimization Libraries (O0/O1) from High-Optimization Libraries (O2/O3) achieves 90.8\% accuracy and 91.0\% precision. This confirms the framework's applicability for identifying libraries that need optimization correction in production apps, where the practical objective is to detect optimization problems rather than pinpoint exact levels.

The full confusion matrix shows that most errors occur between adjacent optimization levels, especially O0/O1 and O1/O2. Full per-ISA breakdowns are included in the artifact.

\textbf{Computational efficiency.} Our classifier processes an average of 847 instruction chunks per second on commodity hardware. The average classification time per library is 0.23 seconds, making the framework suitable for large-scale analysis.

\subsection{Hyperparameter Selection and Stability}
\label{sec:ablation}

We conduct a hyperparameter selection and stability analysis to justify the selected defaults and verify that the main empirical conclusions do not depend on fragile single-point choices. Following controlled one-factor-at-a-time sensitivity protocols used in software analytics and sequence-model ablation studies~\cite{agrawal2019dodge,biedenkapp2017efficient,beltagy2020longformer}, we vary one parameter at a time while fixing all other settings to their defaults. The analysis uses the independent calibration set described above, which is disjoint from the final RQ1 test sets and excluded from the final RQ2/RQ3 prevalence analyses. We evaluate both detection quality (library-level weighted F1) and downstream prevalence stability (low-optimization ratio). For each candidate value $c$ of a parameter and metric $M$, we measure local stability under step perturbations:
\[
\begin{aligned}
\Delta_{\text{mean}}(c)&=\mathbb{E}_{v \in \mathcal{N}(c)}|M(v)-M(c)|,\\
\Delta_{\text{max}}(c)&=\max_{v \in \mathcal{N}(c)}|M(v)-M(c)|.
\end{aligned}
\]
where $\mathcal{N}(c)$ is the admissible local neighborhood produced by adding and subtracting the parameter-specific step. We select the default value that minimizes local variation while preserving near-best validation performance.

\begin{table}[h]
  \centering
  \setlength{\tabcolsep}{3pt}
  \renewcommand{\arraystretch}{0.9}
  \caption{Hyperparameter selection and local stability. Values report mean/max $\Delta$ in percentage points for both F1 and $P$, where lower values indicate greater stability.}
  \label{tab:ablation-study}
  \small
  \resizebox{\columnwidth}{!}{%
  \begin{tabular}{lll}
    \toprule
    \textbf{Parameter} & \textbf{Selected} & \textbf{Other tested candidates} \\
    \midrule
    Window size $W$ &
    \makecell[l]{\textbf{2048}\\F1 0.7/1.4, P 0.8/1.7} &
    \makecell[l]{1024: F1 2.4/4.8, P 2.6/5.1\\3072: F1 1.3/2.6, P 1.5/3.0\\4096: F1 2.1/4.2, P 2.3/4.6} \\
    \midrule
    Stride $S$ &
    \makecell[l]{\textbf{2048}\\F1 0.4/0.9, P 0.6/1.4} &
    \makecell[l]{1024: F1 0.9/1.8, P 1.2/2.5\\1536: F1 0.6/1.2, P 0.8/1.7} \\
    \midrule
    Aggregation weights &
    \makecell[l]{\textbf{(0, 0.25, 0.75, 1, 0.50)}\\F1 0.6/1.3, P 1.1/2.6} &
    \makecell[l]{(0, 0.33, 0.67, 1, 0.50): F1 0.9/1.8, P 1.7/3.9\\(0, 0.20, 0.80, 1, 0.50): F1 1.0/2.1, P 1.9/4.2} \\
    \midrule
    Dominance threshold $\tau_d$ &
    \makecell[l]{\textbf{90\%}\\F1 0.2/0.5, P 0.3/0.8} &
    \makecell[l]{80\%: F1 0.6/1.3, P 0.9/2.0\\85\%: F1 0.3/0.7, P 0.5/1.2\\95\%: F1 0.5/1.1, P 0.8/1.8} \\
    \bottomrule
  \end{tabular}}
\end{table}

Table~\ref{tab:ablation-study} shows that each selected default is the most stable candidate among the tested values. Each cell reports the mean/max absolute change in library-level weighted F1 and low-optimization prevalence ($P$), so smaller values indicate lower sensitivity rather than lower F1 or prevalence. Local neighborhoods use a 512-byte step for $W$ and $S$, a 5-percentage-point step for $\tau_d$, and concrete alternative weighting schemes for aggregation. The aggregation weights are listed in the order O0/O1/O2/O3/Os. For stride, we only compare values no larger than the selected window size ($S \leq W$), because larger strides skip code regions and therefore represent undersampling stress tests rather than ordinary local perturbations.

For window size, $W=2048$ gives the most stable local behavior among the tested values, with lower variation than every alternative on both weighted F1 (0.7/1.4 vs.\ 1.3/2.6 or higher) and low-optimization prevalence (0.8/1.7 vs.\ 1.5/3.0 or higher). This indicates a stable balance between insufficient context (smaller windows) and over-smoothed mixed regions (larger windows). The same pattern holds for stride, aggregation weights, and dominance threshold. Nearby alternatives achieve comparable performance, but the selected defaults yield the smallest local fluctuations in both weighted F1 and low-optimization prevalence. This analysis does not claim a global optimum over all possible parameterizations. It documents the validation logic behind the chosen defaults and shows that the paper's main conclusions are insensitive to reasonable local changes. We therefore keep $W=2048$ for balanced instruction context, $S=2048$ to avoid overlapping-window overhead, the selected aggregation weights to preserve the expected runtime ordering of optimization levels, and $\tau_d=90\%$ to separate nearly pure single-level libraries from genuinely mixed libraries.

\rqbox{
\textbf{Answer to RQ1:} Our optimization detection framework achieves high accuracy (93.0\%) on controlled datasets with single optimization levels and maintains reasonable generalization (81.9\%) on real-world datasets with mixed optimization levels. More importantly, for the core task of distinguishing Low-Optimization Libraries from High-Optimization Libraries, the framework achieves 90.8\% accuracy and 91.0\% precision on real-world data, validating its practical effectiveness for identifying optimization problems in production apps.
}

\section{RQ2: Prevalence of Low Optimization in Real World Apps}
\label{sec:rq2}

To investigate how widespread low compiler optimization practices are in modern mobile software, we conduct a large-scale empirical study on real-world top-ranked mobile apps. This investigation consists of data collection, prevalence analysis, analysis of reused libraries, and case study analysis.
\subsection{Data Collection and Methodology}

We collected 830 available top-ranked apps from Google Play Store (July 2025) across six categories: grossing apps, free apps, grossing games, free games, grossing wearable apps, and free wearable apps. We extracted and analyzed 21,972 native libraries from these apps. During extraction, we applied filtering criteria to exclude libraries unsuitable for optimization analysis: stub libraries (containing only symbol redirects with no executable code), encrypted or packed binaries (where the .text section is obfuscated), and heavily obfuscated binaries where Capstone failed to disassemble more than 50\% of the .text section. These exclusion criteria were applied uniformly across all app categories.

\subsection{Prevalence Analysis}

Table~\ref{tab:opt-distribution} presents the distribution of optimization levels across our dataset. For all prevalence analyses in this paper (RQ2 and RQ3), we define two score-based categories derived directly from the transition zones established in Section~\ref{sec:implementation}: a library is \textit{low-optimization} if its score falls below 0.50 (midpoint of the O1/O2 transition zone), and \textit{high-optimization} if its score reaches 0.80 or above (upper O2/O3 zone). Our analysis reveals that low compiler optimizations are widespread in real-world mobile apps.


\begin{table}[h]
\centering
\setlength{\tabcolsep}{4pt}
\renewcommand{\arraystretch}{0.9}
\caption{Distribution of Optimization Levels Across Application Categories (Including Wearables)}
  \label{tab:opt-distribution}
  \small
  \resizebox{\columnwidth}{!}{%
    \begin{tabular}{lcccccc}
    \toprule
    \textbf{Category} & \textbf{.so} & \textbf{Avg. Opt.} & \textbf{Low Opt.} & \textbf{Low Opt.} & \textbf{High Opt.} & \textbf{High Opt.} \\
    & \textbf{Files} & \textbf{Score} & \textbf{Files} & \textbf{Ratio} & \textbf{Files} & \textbf{Ratio} \\
    \midrule
      Grossing Apps     & 5,100 & 0.565 & 1,238 & 24.3\% & 243 & 4.8\% \\
      Top Free Apps     & 4,566 & 0.548 & 1,299 & 28.4\% & 242 & 5.3\% \\
      Grossing Games    & 3,084 & 0.520 & 1,009 & 32.7\% & 125 & 4.1\% \\
      Top Free Games    & 3,884 & 0.510 & 1,416 & 36.5\% &  86 & 2.2\% \\
      Grossing Wears    & 2,592 & 0.525 &   809 & 31.2\% &  96 & 3.7\% \\
      Top Free Wears    & 2,746 & 0.511 &   938 & 34.2\% & 103 & 3.8\% \\
      \midrule
      \textbf{Total}    & \textbf{21,972} & \textbf{0.534} & \textbf{6,709} & \textbf{30.5\%} & \textbf{895} & \textbf{4.1\%} \\
      \bottomrule
    \end{tabular}%
  }
\end{table}

The results demonstrate significant variation across application categories. Among regular apps, both grossing apps and free apps show similar optimization patterns, with 24.3\% and 28.4\% of .so files exhibiting low optimization scores and average optimization scores of 0.565 and 0.548, respectively. Notably, high-optimization libraries are rare across all categories, with only 4.1\% of total libraries classified as high-optimization (score $\geq$ 0.80). Mobile games demonstrate markedly worse optimization practices, with average optimization scores of 0.510--0.520 and 32.7\%--36.5\% of .so files classified as low-optimization.

Wearable applications exhibit intermediate optimization patterns, with grossing wears and free wears showing 31.2\% and 34.2\% low-optimization ratios respectively, and average optimization scores of 0.525 and 0.511. This positions wearable apps between regular apps and games in terms of optimization quality, confirming a consistent trend across all app types.

Overall, 30.5\% of all analyzed native libraries use low optimization levels, affecting 91.7\% of apps, indicating that optimization issues are systemic rather than isolated. This app-level figure is not driven by isolated single-library cases. Specifically, 85.8\% of apps have more than 10\% low-optimization libraries, 58.9\% have more than 25\%, and 10.6\% have more than 50\%. The median per-app low-optimization ratio is 28.8\%.

\textbf{Threshold robustness.} We further test whether the prevalence result is sensitive to the low-optimization cutoff. Using three cutoffs around the O1/O2 transition zone, the low-optimization prevalence is 21.3\% at 0.35, 30.5\% at our midpoint cutoff 0.50, and 58.5\% at 0.65. The 0.35 cutoff captures only the most severe cases, while 0.65 over-flags libraries in the O2 range and changes the category ranking. We therefore use 0.50 as the balanced point where low- and high-optimization chunk contributions meet. For the high-optimization cutoff, the rate drops from 24.5\% at 0.70 to 4.1\% at 0.80, and then only to 0.5\% at 0.90, making 0.80 the natural knee point. The linear weighting scheme could in principle mask heavily O0-contaminated mixed libraries, but empirically this boundary case is rare: about 70\% of low-optimization libraries score below 0.35 (36.6\% in $[0,0.15)$ and 33.2\% in $[0.15,0.35)$), and only 30.0\% fall in $[0.35,0.50)$. Among libraries with score $\geq$ 0.50, only 0.08\% contain more than 50\% O0/O1 chunks.

The disparity between categories is notable. Games typically demand high performance yet show the worst optimization practices (32.7\%--36.5\% low-optimization), while wearable apps with constrained hardware resources perform better than games but worse than regular apps (31.2\%--34.2\% vs.\ 24.3\%--28.4\%).

\textbf{Detailed Optimization Level Analysis.} To gain deeper insights into optimization practices, we analyze the distribution of optimization levels at the code chunk granularity. Table~\ref{tab:opt-chunks-distribution} presents the detailed breakdown of optimization level chunks across application categories.

\begin{table}[h]
\centering
\setlength{\tabcolsep}{4pt}
\renewcommand{\arraystretch}{0.9}
\caption{Distribution of Optimization Level Chunks Across Application Categories (Including Wearables). Chunks represent code segments identified by our classifier.}
\label{tab:opt-chunks-distribution}
\small
\resizebox{\linewidth}{!}{%
\begin{tabular}{lcccccc}
\toprule
\textbf{Category} & 
\makecell{\textbf{O0}\\\textbf{Chunks}} & 
\makecell{\textbf{O1}\\\textbf{Chunks}} & 
\makecell{\textbf{O2}\\\textbf{Chunks}} & 
\makecell{\textbf{O3}\\\textbf{Chunks}} &
\makecell{\textbf{Os}\\\textbf{Chunks}} &
\makecell{\textbf{Total}\\\textbf{Chunks}} \\
\midrule
Grossing Apps    & \makecell{404,919\\{\small(16.1\%)}} & \makecell{216,900\\{\small(8.6\%)}}  & \makecell{726,524\\{\small(29.0\%)}} & \makecell{490,291\\{\small(19.5\%)}} & \makecell{669,277\\{\small(26.7\%)}} & 2,507,911 \\[8pt]

Top Free Apps    & \makecell{285,563\\{\small(15.3\%)}} & \makecell{175,680\\{\small(9.4\%)}}  & \makecell{578,810\\{\small(31.0\%)}} & \makecell{379,772\\{\small(20.3\%)}} & \makecell{448,620\\{\small(24.0\%)}} & 1,868,445 \\[8pt]

Grossing Games   & \makecell{794,023\\{\small(27.1\%)}} & \makecell{189,877\\{\small(6.5\%)}}  & \makecell{825,067\\{\small(28.1\%)}} & \makecell{621,328\\{\small(21.2\%)}} & \makecell{503,977\\{\small(17.2\%)}} & 2,934,272 \\[8pt]

Top Free Games   & \makecell{671,109\\{\small(29.5\%)}} & \makecell{151,955\\{\small(6.7\%)}}  & \makecell{620,904\\{\small(27.3\%)}} & \makecell{433,099\\{\small(19.0\%)}} & \makecell{399,799\\{\small(17.6\%)}} & 2,276,866 \\[8pt]

Grossing Wears   & \makecell{165,915\\{\small(18.3\%)}} & \makecell{94,710\\{\small(10.5\%)}}  & \makecell{268,596\\{\small(29.7\%)}} & \makecell{153,181\\{\small(16.9\%)}} & \makecell{222,409\\{\small(24.6\%)}} & 904,811 \\[8pt]

Top Free Wears   & \makecell{205,666\\{\small(18.1\%)}} & \makecell{108,495\\{\small(9.6\%)}}  & \makecell{348,875\\{\small(30.8\%)}} & \makecell{219,941\\{\small(19.4\%)}} & \makecell{250,191\\{\small(22.1\%)}} & 1,133,168 \\[8pt]

\midrule
\textbf{Total}   & \makecell{\textbf{2,527,195}\\{\small\textbf{(21.7\%)}}} & \makecell{\textbf{937,617}\\{\small\textbf{(8.1\%)}}} & \makecell{\textbf{3,368,776}\\{\small\textbf{(29.0\%)}}} & \makecell{\textbf{2,297,612}\\{\small\textbf{(19.8\%)}}} & \makecell{\textbf{2,494,273}\\{\small\textbf{(21.5\%)}}} & \textbf{11,625,473} \\
\bottomrule
\end{tabular}%
}

\end{table}

The chunk-level analysis shows consistent patterns across categories. Across all 11,625,473 analyzed chunks, O0 accounts for 21.7\% and O1 for only 8.1\%, while high-optimization chunks (O2+O3) total 48.7\% and Os accounts for 21.5\%. At the category level, games have the worst distribution: grossing games contain 27.1\% O0 chunks and 49.3\% high-optimization (O2+O3) chunks, compared to regular apps at 15.3--16.1\% O0 and 48.5--51.3\% high-optimization. Free games show slightly worse values at 29.5\% O0 and 46.3\% high-optimization. Wearable applications fall in between at 18.1--18.3\% O0 and 46.6--50.2\% high-optimization chunks. Os (size-optimized) chunk prevalence varies from 17.2\% in grossing games to 26.7\% in grossing apps, with wearables at 22.1--24.6\%. Notably, O1 chunks are consistently the smallest proportion across all categories (6.5\%--10.5\%). This suggests that libraries tend to be either debug builds (O0) inadvertently shipped in release packages, or properly optimized release builds (O2/O3). O1 is rarely a deliberate choice in practice.

\subsection{Analysis of Reused Libraries}

Table~\ref{tab:reused-libs-real} presents the most frequently reused native libraries that consistently exhibit low optimization scores (average score below 0.50), each appearing in at least 10 apps.

\begin{table}[h]
  \centering
  \setlength{\tabcolsep}{4pt}
  \renewcommand{\arraystretch}{0.9}
  \caption{Most Common Low-Optimization Native Libraries}
  \label{tab:reused-libs-real}
  \resizebox{\columnwidth}{!}{%
    \begin{tabular}{lcccc}
    \toprule
    \textbf{Library} & \textbf{Occs.} & \textbf{Avg. Opt.} & \textbf{Low Opt.} & \textbf{Example} \\
    \textbf{Name} & & \textbf{Score} & \textbf{Ratio} & \textbf{Apps} \\
    \midrule
      libcrashlytics-trampoline.so & 146 & 0.023 & 96.6\% & Firebase Crashlytics, Unity Games \\
      libmain.so & 145 & 0.218 & 61.4\% & Various Unity Apps, Native Games \\
      libil2cpp.so & 141 & 0.317 & 80.9\% & Unity Games, C\# Mobile Apps \\
      libtobEmbedPagEncrypt.so & 95 & 0.427 & 72.6\% & Chinese Apps, Security SDKs \\
      libdatastore\_shared\_counter.so & 71 & 0.218 & 100.0\% & Google Play Services, Analytics \\
      libbuffer.so & 57 & 0.363 & 94.7\% & Media Apps, Buffer Management \\
      libsentry-android.so & 41 & 0.351 & 85.4\% & Error Tracking, Crash Reporting \\
      libsurface\_util\_jni.so & 36 & 0.258 & 91.7\% & Graphics Apps, Surface Rendering \\
      libbugsnag-root-detection.so & 28 & 0.071 & 92.9\% & Security Apps, Root Detection \\
      libsqlite3.so & 27 & 0.496 & 33.3\% & Database Apps, Local Storage \\
      \bottomrule
    \end{tabular}%
  }
\end{table}

The most concerning case is \texttt{libcrashlytics-\allowbreak trampoline.so}, which appears in 146 apps with an average optimization score of only 0.023. This case exemplifies the widespread impact of third-party SDK optimization issues, as crash reporting is critical for application stability yet performs poorly due to low-optimization compilation.

Similarly problematic is \texttt{libil2cpp.so}, Unity's IL2CPP runtime library found in 141 applications (primarily games) with an optimization score of 0.317. Given Unity's dominance in mobile game development, this represents a significant performance bottleneck affecting millions of users.

These findings reveal two primary sources of optimization problems: (1) third-party SDKs distributed as precompiled binaries with poor optimization settings, and (2) widely-used development frameworks (e.g., Unity) that may not prioritize optimization in their default build configurations.
\rqbox{
\textbf{Answer to RQ2:} Low compiler optimizations are widespread, affecting 30.5\% of 21,972 analyzed native libraries across 830 apps, with 91.7\% of apps containing at least one low-optimization library and a median per-app low-optimization ratio of 28.8\%. Only 4.1\% of libraries achieve high optimization levels. Games show the worst optimization practices (32.7\%--36.5\% low-optimization), followed by wearable apps (31.2\%--34.2\%), with regular apps performing best (24.3\%--28.4\%). Frequently reused third-party SDKs and development framework libraries are a primary source of the problem.
}

\section{RQ3: Multi-dimensional Impacts of Optimization Fixes}

To understand the real-world impacts of addressing compiler optimization problems, we investigate \textbf{RQ3: What are the multi-dimensional impacts of fixing optimization problems in production apps?} This research question examines three dimensions: (1) technical performance improvements, (2) user-perceived quality improvements, and (3) the underlying sources of optimization problems. We conduct an in-depth analysis of 12 production apps: 6 top-ranked commercial apps and 6 open-source apps published on Google Play Store.

\subsection{Research Design and Methodology}

\textbf{Case Selection.}
We select 12 production apps: 6 top-ranked commercial apps and 6 open-source apps. Selection criteria include app popularity (millions of active users), presence of native libraries with identified optimization issues, and feasibility of collaboration. Commercial apps include Payment App A (payment platform), Video App B (video streaming), Social App C (social media), Card Game X (strategy game), FPS Game B (first-person shooter), and MOBA Game C (multiplayer battle arena), anonymized per confidentiality agreements. Open-source apps additionally require accessible source code and complete build systems: VLC, Kodi, Firefox, Termux, Signal, and Telegram.

\begin{table}[h]
  \centering
  \setlength{\tabcolsep}{4pt}
  \renewcommand{\arraystretch}{0.9}
  \caption{Case Study Apps: Commercial and Open-Source}
  \label{tab:partner-apps}
  \small
  \resizebox{\columnwidth}{!}{%
  \begin{tabular}{p{2cm}p{3.8cm}|p{1.5cm}p{4cm}}
    \toprule
    \multicolumn{2}{c}{\textbf{Commercial Apps (Case Set 1)}} & \multicolumn{2}{c}{\textbf{Open-Source Apps (Case Set 2)}} \\
    \cmidrule(lr){1-2} \cmidrule(lr){3-4}
    \textbf{App} & \textbf{Description} & \textbf{App} & \textbf{Description} \\
    \midrule
    Payment App A & Leading mobile payment platform with QR code scanning & VLC & Multimedia player with FFmpeg codecs (2.5K+ stars, 100M+ downloads) \\
    Video App B & Video streaming and content creation platform & Kodi & Media center with native rendering (16K+ stars, 50M+ downloads) \\
    Social App C & Social media and content sharing app & Firefox & Web browser with Gecko engine (1.2K+ stars, 500M+ downloads) \\
    Card Game X & Strategy card game with complex rendering & Termux & Terminal emulator with native execution (25K+ stars, 10M+ downloads) \\
    FPS Game B & First-person shooter with intensive graphics & Signal & Encrypted messaging with crypto libraries (42K+ stars, 100M+ downloads) \\
    MOBA Game C & Multiplayer battle arena using Unity3D & Telegram & Messaging app with native libraries (25K+ stars, 1B+ downloads) \\
    \bottomrule
  \end{tabular}}
\end{table}

\subsection{RQ3.1: Technical Performance Improvements}

\textbf{Methodology.}
We measure technical performance improvements achieved through optimization correction using the six commercial apps (Case Set 1). These apps enable controlled experiments with precise deployment dates and production validation. For each app, we: (1) identify optimization issues using our detection framework, (2) recompile affected libraries with appropriate optimization levels (O0/O1 $\to$ O2/O3), (3) measure performance improvements using controlled experiments, and (4) validate correctness and security properties. Correctness validation is performed by partner development teams using their existing test suites (unit, integration, and functional tests), with particular attention to floating-point numerical stability (comparing outputs within tolerance bounds), timing-sensitive operations, and cryptographic routines where constant-time execution is required. All optimized libraries passed their respective test suites before production deployment.

\textbf{Performance Metrics.}
We use CPU instruction count as our primary runtime performance indicator, as it directly quantifies computational work and strongly correlates with execution time and energy consumption~\cite{herglotz2023video,georgiou2021comprehensive}. Specifically, we measure \textit{retired instructions} via hardware Performance Monitoring Unit (PMU) counters, which count only instructions that complete execution and produce committed results, excluding speculative or flushed instructions. While absolute instruction counts are device-specific, the \textit{relative reduction} between the original (O0/O1) and optimized (O2/O3) builds is always measured on the \textbf{same device under identical workloads}, so hardware variation cancels out and the percentage reduction is a device-controlled comparative metric that reflects only the optimization-level change. This approach is widely used in compiler optimization research~\cite{muchnick1997advanced}.

\textbf{Measurement Protocol.}
We establish baselines using original binaries with identified optimization issues. We or our partner development teams then recompile affected libraries with high optimization levels (O2 for general code, O3 for computationally intensive components) while preserving all other build settings. We execute identical test scenarios on both versions using automated test scripts to ensure reproducibility. We average results over multiple measurement iterations (minimum 5 runs per configuration, maximum 10 runs when variance is high) to ensure statistical validity. Standard deviations across runs were below 1.5\% of the mean for all reported instruction-count reductions, confirming measurement stability. We collect PMU-based retired instruction counts on a Xiaomi 14 Android device (Snapdragon 8 Gen 3), using Android simpleperf/perf\_event-based profiling and Perfetto/Android Studio for trace inspection.

\textbf{Results.}
Table~\ref{tab:rq3-commercial-results} summarizes results for all six commercial apps. CPU instruction reductions range from 10\% to 63\% (median: 20.5\%). The variation reflects two factors: the number of libraries successfully optimized and their execution frequency, as libraries on the critical path yield the largest gains. Payment App A contained 98 unoptimized libraries, of which 6 were successfully optimized under collaboration constraints, while other apps had more focused interventions (1--10 libraries). Payment App A achieves 22\% reduction. Card Game X achieves 63\% reduction with 40\% power consumption reduction and 15 FPS gain. Video App B achieves 25\% reduction with 15\% streaming latency reduction and 30\% frame-drop reduction. Social App C achieves 19\% reduction across 8 of 17 libraries with improvements in feed scrolling and image loading. FPS Game B achieves 10\% reduction with thermal throttling mitigation. MOBA Game C achieves 13\% reduction alongside 65\% binary size reduction (37MB$\to$13MB) and 35\% load-time reduction.
These cases show that the impact of optimization fixes is driven less by the number of recompiled libraries than by whether the affected libraries lie on latency-critical execution paths.

\begin{table}[h]
  \centering
  \setlength{\tabcolsep}{4pt}
  \renewcommand{\arraystretch}{0.9}
  \caption{Performance Improvements for Commercial Apps (Case Set 1). All apps deployed to production.}
  \label{tab:rq3-commercial-results}
  \small
  \resizebox{0.94\columnwidth}{!}{%
  \begin{tabular}{p{1.8cm}p{4cm}p{3.5cm}c}
      \toprule
      \textbf{App} &
      \makecell{\textbf{Optimized Libraries}\\\textbf{(Fixed/Total)}} &
      \makecell{\textbf{Additional Metrics}\\\textbf{(Disclosed)}} &
      \textbf{CPU Reduc.} \\
      \midrule
    Payment App A & \makecell[l]{libqrscanner.so, libimageproc.so\\(6/98)} & \makecell[l]{Cold start: -200ms\\QR init: -60\%, Binary: -22\%} & 22\% \\
      \midrule
    Video App B & \makecell[l]{libbroadcast-client.so, libstream.so\\(2/3)} & \makecell[l]{Streaming: -15\% latency\\Frame drops: -30\%} & 25\% \\
      \midrule
    Social App C & \makecell[l]{libsocial-log.so, libnetwork.so, ...\\(8/17)} & \makecell[l]{Feed scrolling: smoother\\Image loading: faster} & 19\% \\
      \midrule
    Card Game X & \makecell[l]{librender.so, libgame-engine.so\\(2/2)} & \makecell[l]{Power: -40\%\\Frame rate: +15 FPS} & 63\% \\
      \midrule
    FPS Game B & \makecell[l]{10 O0 libs (graphics, physics, audio)\\(10/10)} & \makecell[l]{Latency: -8\%\\Thermal throttling: reduced} & 10\% \\
      \midrule
    MOBA Game C & \makecell[l]{il2cpp.so, Unity 3D\\(1/1)} & \makecell[l]{Binary: -65\% (37$\to$13MB)\\Load time: -35\%} & 13\% \\
      \bottomrule
  \end{tabular}}
\end{table}

\textbf{Case Study: Payment App A.}
Payment App A (100M+ QR scans/day) had persistent user complaints about slow QR scanning despite months of manual code optimization by the development team. \textsc{OptDetect} identified that 6 QR-module libraries (libqrscanner.so, libimageproc.so, libcamera-util.so, etc.) were compiled at O0. Investigation revealed that app development, maintenance, and release were handled by separate teams, and during one release cycle a release team had accidentally packaged debug-version libraries into the production app, a mistake that persisted undetected across multiple releases. Recompiling at O2/O3 achieved 22\% CPU instruction reduction, 60\% camera initialization speedup (800ms$\to$320ms), 200ms cold-start improvement, and 22\% binary size reduction (321MB$\to$249MB).
This case illustrates that low-optimization binaries can persist even in mature release pipelines when development, maintenance, and packaging responsibilities are separated.

\subsection{RQ3.2: User-Perceived Quality Improvements}

We analyze app store reviews to test whether technical improvements translate to user-perceived quality changes. We collect 13,156 reviews from multiple app stores across all of 2025 for the 6 commercial apps. Reviews are filtered to the five most frequent languages (English, Simplified Chinese, Traditional Chinese, Korean, Japanese), covering over 92\% of all reviews. Performance-related keywords cover four categories with equivalent terms in all five languages: \textit{performance symptoms} (e.g., ``lag'', ``freeze''), \textit{resource complaints} (e.g., ``battery drain'', ``overheat''), \textit{loading issues} (e.g., ``slow startup''), and \textit{rendering issues} (e.g., ``frame drop'', ``jank'').

We compare monthly keyword frequency and average rating for the optimization month vs.\ the following month. A one-tailed Wilcoxon signed-rank test across the six apps yields $p = 0.031$ ($\alpha = 0.05$), confirming statistically significant keyword reduction. Note that keyword reduction is associated with, but not solely caused by, the optimization fix, as concurrent app updates may also contribute. Potential confounds are discussed in Threats to Validity.

\begin{figure*}[t]
  \centering
  \includegraphics[width=0.56\textwidth]{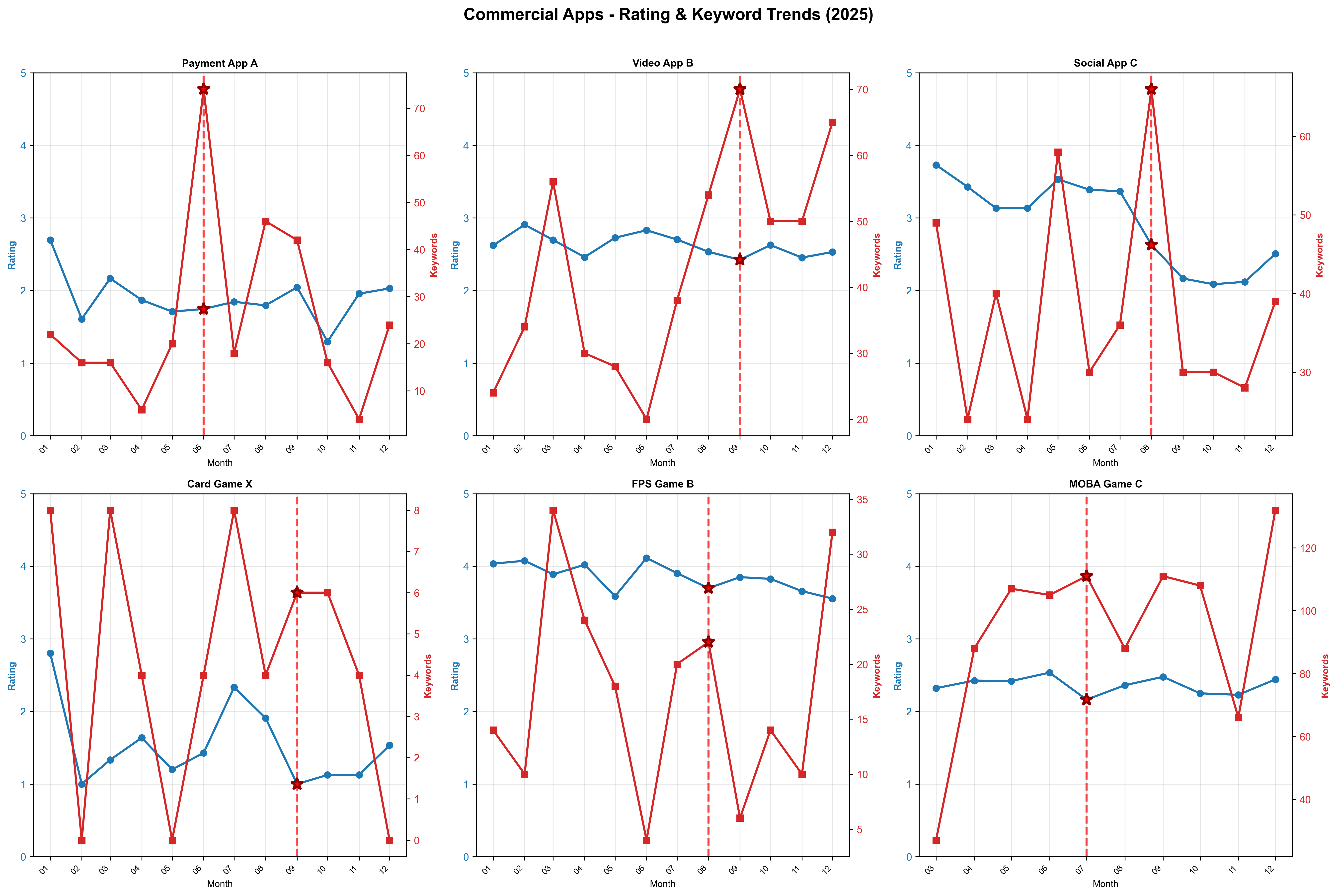}
  \caption{Monthly rating (blue, left y-axis) and performance-related keyword frequency (red, right y-axis) trends for six commercial apps throughout 2025. Red dashed lines and star markers indicate .so optimization intervention dates. Five apps show immediate keyword reductions in the month following optimization, while Card Game X shows delayed but sustained decline over three follow-up months (6$\to$6$\to$4$\to$0).}
  \Description{A multi-panel trend figure showing monthly app ratings and performance-related keyword frequencies for six commercial apps, with markers indicating optimization intervention dates.}
  \label{fig:commercial-apps-2025-trends}
\end{figure*}

\textbf{Results.}
Figure~\ref{fig:commercial-apps-2025-trends} and Table~\ref{tab:rq3-commercial-perception} show the results. Performance-related keyword frequency decreases in 5 of 6 apps (21\%--76\%), with a median change of --42\% across all six apps. Ratings improve in 5 of 6 apps (+0.09 to +0.21 stars), with a median change of +0.14 across all six apps.

\textbf{Payment App A} achieves the most dramatic improvement (--76\% keyword reduction, from 74 to 18), directly corresponding to its 22\% CPU instruction reduction and 60\% QR initialization speedup. \textbf{FPS Game B} shows similarly strong results (--73\%, from 22 to 6) despite the lowest CPU reduction (10\%), suggesting that even modest improvements on user-interactive features translate to substantial user perception gains. \textbf{Card Game X} shows a delayed but sustained decline (6$\to$6$\to$4$\to$0 over three follow-up months), ultimately eliminating all performance-related complaints after the largest CPU reduction (63\%) in the set takes full effect. \textbf{Video App B} shows balanced improvement (--29\% keywords, +0.21 stars), with the highest rating increase among all apps. \textbf{MOBA Game C} achieves a moderate keyword reduction (--21\%) despite the largest binary size reduction (65\%), with the persistently high keyword volume reflecting the competitive gaming context where users are highly sensitive to any performance variation.

\textbf{Social App C} is the only exception, showing keyword reduction (--55\%) but a rating drop (--0.46 stars). Because aggregate star ratings are influenced by many factors beyond performance, we consider performance-symptom keyword frequency the more precise signal for evaluating optimization impact.

\begin{table}[h]
  \centering
  \setlength{\tabcolsep}{4pt}
  \renewcommand{\arraystretch}{0.9}
  \caption{User Perception Changes for Commercial Apps After .so Optimization (2025 Data). (Opt$\to$Next) indicates comparison between the optimization deployment month and the following month.}
  \label{tab:rq3-commercial-perception}
  \small
  \resizebox{\columnwidth}{!}{%
  \begin{tabular}{lccccc}
      \toprule
      \textbf{App} &
      \textbf{Optimization} &
      \textbf{Keywords} &
      \textbf{Keyword} &
      \textbf{Rating} &
      \textbf{Rating} \\
      & \textbf{Date} & \textbf{(Opt$\to$Next)} & \textbf{Change} & \textbf{(Opt$\to$Next)} & \textbf{Change} \\
      \midrule
    Payment App A & 2025-06 & 74$\to$18 & -76\% & 1.75$\to$1.84 & +0.09 \\
    Video App B & 2025-09 & 70$\to$50 & -29\% & 2.42$\to$2.63 & +0.21 \\
    Social App C & 2025-08 & 66$\to$30 & -55\% & 2.63$\to$2.17 & -0.46 \\
    Card Game X & 2025-09 & 6$\to$6$\to$4$\to$0 & 0\% & 1.00$\to$1.12 & +0.12 \\
    FPS Game B & 2025-08 & 22$\to$6 & -73\% & 3.70$\to$3.85 & +0.15 \\
    MOBA Game C & 2025-07 & 111$\to$88 & -21\% & 2.16$\to$2.36 & +0.20 \\
      \midrule
    \textbf{Median} & - & - & \textbf{-42\%} & - & \textbf{+0.14} \\
      \bottomrule
  \end{tabular}}
\end{table}

\subsection{Open-Source App Results and Root Cause Analysis}

Unlike commercial apps where source code is inaccessible, open-source apps allow us to trace each low-optimization library to its origin through build script and source dependency analysis. For each of the six open-source apps, we examine whether low-optimization libraries are compiled from the project's own source code or included as third-party pre-compiled binaries.

Across 314 native libraries in these six apps, we identify 25 low-optimization libraries (8.0\%). Table~\ref{tab:rq3-opensource-results} presents the per-app tracing results. The findings are clear: 24 out of 25 libraries (96.0\%) originate from third-party pre-compiled binaries distributed by external vendors or upstream repositories. The single exception is Telegram's libtmessages.49.so, which is traced to an app-level build configuration error. Kodi alone accounts for 15 low-optimization libraries, all from the PyCryptodome Python crypto extension shipped as pre-compiled .so modules. Using the same measurement protocol as RQ3.1, CPU instruction reductions after fixing these libraries range from 15\% to 58\% (median: 32\%).

\begin{table}[h]
  \centering
  \setlength{\tabcolsep}{4pt}
  \renewcommand{\arraystretch}{0.9}
  \caption{Root Cause Tracing for Open-Source Apps (Case Set 2). Each low-optimization library is traced to its origin via build script and source dependency analysis.}
  \label{tab:rq3-opensource-results}
  \small
  \resizebox{\columnwidth}{!}{%
  \begin{tabular}{p{1.3cm}cp{3.2cm}p{3cm}c}
      \toprule
      \textbf{App} &
      \makecell{\textbf{Low-Opt}\\\textbf{Count}} &
      \textbf{Identified Libraries} &
      \textbf{Traced Origin} &
      \textbf{CPU Reduc.} \\
      \midrule
    VLC & 1 & libvlcjni.so & Third-party (FFmpeg binding) & 35\% \\
      \midrule
    Firefox & 2 & libclearkey.so, libsoftokn3.so & Third-party (NSS crypto) & 23\% \\
      \midrule
    Signal & 4 & libsqlcipher.so, libaesgcm.so, libimage\_processing\_util\_jni.so ($\times$2: x86/x86\_64) & Third-party (crypto and media processing) & 28\% \\
      \midrule
    Termux & 2 & libtermux-bootstrap.so, libproot-loader.so & Third-party (system tools) & 42\% \\
      \midrule
    Kodi & 15 & PyCryptodome modules (15 .so files) & Third-party (Python crypto extension) & 58\% \\
      \midrule
    Telegram & 1 & libtmessages.49.so & Build config error & 15\% \\
      \bottomrule
  \end{tabular}}
\end{table}

\textbf{Developer Interviews.}
From the development and maintenance teams of all 12 apps, we recruit 18 developers who meet our criteria ($\geq$3 years of mobile development experience with direct knowledge of build systems and release pipelines) and agree to participate in semi-structured interviews. Based on their responses, we identify four systemic challenges that contribute to low-optimization libraries persisting in production. (1) \textit{SDK vendor practices}: vendors often ship debug builds to facilitate crash debugging, and requesting optimized builds takes a median of 3--6 months when successful. (2) \textit{Dependency chain complexity}: transitive dependencies are largely invisible to app developers, making it difficult to audit optimization levels across the full dependency tree. (3) \textit{Process issues}: debug builds are accidentally released to production due to separated development and release teams. (4) \textit{Lack of detection tools}: prior to this work, no tooling existed to identify optimization problems in production app binaries, so the issue went undetected regardless of severity.
Together, these responses indicate that the issue is not merely a local compiler-flag mistake, but a supply-chain visibility problem spanning SDK vendors, app developers, and repository maintainers.

\subsection{Ecosystem-Level Investigation}
The interview findings above indicate that developers lack both the tools and the process visibility to audit optimization levels of third-party libraries. In practice, given the complexity of modern app build pipelines and the absence of detection tooling, developers have no choice but to assume that libraries from official repositories are properly optimized. To investigate whether this assumption holds, we conduct an ecosystem-level analysis of one such major repository.

We investigate the official third-party library repository of a major mobile platform, which is the primary distribution channel for native libraries used by apps on that platform. Per our confidentiality agreement with the repository maintainers, we anonymize the repository identity. We analyze all 1,073 native libraries available in this repository as of December 2025. Applying the same score threshold defined in Section~\ref{sec:rq2} (score $<$ 0.50), our analysis reveals that 49.7\% of libraries (533 out of 1,073) exhibit low optimization levels. The majority are classified as O0-dominant (467) or O1-dominant (35) by the tool's dominance criterion ($>$90\% of chunks at that level), indicating nearly pure low-optimization compilation. The remaining 31 have mixed-level scores in the O1/O2 transition zone (0.35--0.50). The snapshot is included in our artifact for reproducibility. Table~\ref{tab:ecosystem-function-distribution} presents the functional breakdown of the 502 O0- and O1-dominant libraries.

\begin{table}[h]
  \centering
  \setlength{\tabcolsep}{4pt}
  \renewcommand{\arraystretch}{0.9}
  \caption{Functional Distribution of O0/O1-Dominant Low-Optimization Libraries in Repository}
  \label{tab:ecosystem-function-distribution}
  \small
  \begin{tabular}{lcc}
    \toprule
    \textbf{Function Category} & \textbf{Count} & \textbf{\% of Dominant Low-Opt} \\
    \midrule
    System Utilities \& FFI & 420 & 83.7\% \\
    Multimedia/Codec & 26 & 5.2\% \\
    Cryptography & 22 & 4.4\% \\
    Networking & 16 & 3.2\% \\
    Location/Map & 7 & 1.4\% \\
    Image Processing & 5 & 1.0\% \\
    Data Compression & 5 & 1.0\% \\
    Speech/Audio & 1 & 0.2\% \\
    \midrule
    \textbf{Total} & \textbf{502} & \textbf{100\%} \\
    \bottomrule
  \end{tabular}
\end{table}

System utilities and FFI interfaces constitute the majority (83.7\%, 420 libraries), including general-purpose utilities (libentry.so, liblibrary.so), logging frameworks (libaliyunlog.so, 5.2 MB), location services (liblocsdk8b.so), and HTTP servers (libmongoose.so). While individual utilities may not be hotspots, the cumulative effect of numerous low-optimization libraries is significant as modern apps integrate dozens of them.

Multimedia codecs rank second (26 libraries, 5.2\%), including large video players (libHJPlayer.so, 23.41 MB) and FFmpeg-based decoders (libwlffmpeg.so, 18.57 MB). This is particularly concerning because multimedia processing is a performance-critical hotspot in mobile applications. Poorly optimized codecs directly impact video playback, audio quality, and media loading, making their optimization gaps more severe in practice. Affected libraries span a wide size range from 3.96 KB to 25.39 MB (average: 712.76 KB, median: 4.64 KB), with the largest files representing the most severe bottlenecks due to high instruction overhead and memory footprint. Cryptography libraries (4.4\%) and networking libraries (3.2\%) are also noteworthy, as both are frequently invoked during core operations such as authentication, data encryption, and API communication. Low optimization in these categories can degrade responsiveness in latency-sensitive user workflows.

\textbf{Repository Implication.}
These results suggest that repository-level screening should prioritize large and performance-sensitive native libraries rather than treating all binaries equally. A lightweight batch scan with \textsc{OptDetect} would have flagged dominant O0/O1 libraries and provided maintainers with actionable evidence before such artifacts propagated into downstream apps.

\textbf{Repository Acknowledgement.}
We disclosed our findings to the repository maintainers. They explained that most libraries are submitted as pre-compiled binaries, and the repository previously had no batch-level mechanism to verify their optimization quality. They also noted that submitted artifacts were typically treated as production-ready, making such issues difficult to detect without an automated tool. After reviewing our detection evidence, the maintainers formally acknowledged the presence of many debug-level compilation configurations and are working with affected library developers to address the issue.

\rqbox{
\textbf{Answer to RQ3:} Fixing compiler optimization problems in production apps yields measurable impacts. CPU instruction reductions range from 10\%--63\% (median: 20.5\%) for 6 commercial apps and 15\%--58\% (median: 32\%) for 6 open-source apps. Performance-symptom keyword frequency decreases in 5 of 6 commercial apps, with a median change of --42\% across all six apps ($p = 0.031$), providing a more targeted signal of optimization impact than aggregate star ratings. Root cause tracing in open-source apps shows that 96\% (24/25) of low-optimization libraries originate from third-party pre-compiled binaries. An ecosystem-level investigation further confirms the systemic nature of this problem: 49.7\% of 1,073 libraries exhibit low optimization levels.
}

\section{Threats to Validity}

\textbf{Confidence Score Usage.} Our framework computes an entropy-based confidence score for each library classification and reports it as a supplementary diagnostic output rather than using it for automated filtering. Confidence-weighted prevalence estimation is left to future work.

\textbf{Framework Generalizability.} Our framework's accuracy depends on training data representativeness. Section~\ref{sec:ablation} reports local sensitivity ablations for the main empirical parameters, while broader architecture searches remain future work. We adopt a unified cross-ISA model because recurring optimization patterns can generalize across architectures and a unified model benefits from a larger combined training set, while per-ISA ablations remain future work. We use BiLSTM due to its strong sequence modeling and low training cost, and alternative architectures such as CNNs or Transformers remain future work. The aggregation weights and prevalence thresholds were empirically tuned on the independent calibration set described in RQ1. Although we use both Optimization-Detector and Assemblage, they may not cover all compiler versions and build configurations. Our framework may also face challenges with heavily obfuscated or packed binaries that obscure optimization patterns.

\textbf{Dataset Sampling.} Our dataset comprises 830 top-ranked apps from Google Play Store in July 2025. Top-ranked apps may have better optimization practices than average apps, potentially underestimating the ecosystem-wide problem. However, these apps collectively account for the vast majority of user installations and interactions, making optimization issues in this set highly relevant for real-world user impact.

\textbf{User Perception Analysis.} Our user perception analysis uses correlational evidence from app store reviews. Keyword selection may not capture all performance-related feedback, and confounding factors such as new feature releases, bug fixes, or public events may influence rating changes independently of optimization fixes.

\textbf{Metric Limitations.} We use CPU instruction count as our primary performance metric, which provides a reliable and scalable proxy but does not capture all performance dimensions such as cache behavior and memory bandwidth. Additionally, our RQ2 prevalence analysis counts all libraries and chunks statically without weighting by runtime hotness: a low-optimization library that is rarely invoked contributes less actual performance overhead than one on the critical path. While RQ3 demonstrates real-world impact through deployment case studies, the static counts in RQ2 may overestimate ecosystem-wide performance waste for libraries that are infrequently called. Additionally, standard compiler optimization transitions (O0/O1 to O2/O3) may introduce edge-case risks in floating-point rounding, timing-sensitive code, or cryptographic routines requiring constant-time execution. Our validation across 12 apps mitigates but does not eliminate these concerns.

\section{Related Work}

\textbf{Compiler Optimization Detection and Binary Analysis.}
Research on compiler optimization detection has primarily focused on desktop and server environments. Previous work has explored optimization level identification through static analysis of assembly code patterns~\cite{banerjee2014detecting}, performance counter analysis~\cite{chowdhury2019greenoracle}, and compiler fingerprinting techniques. Recent advances include binary-level optimization detection using deep learning~\cite{duan2020deepbindiff, ren2021unleashing} and transparent compiler optimization frameworks~\cite{mpeis2021developer}.

Binary analysis techniques have been extensively developed for malware detection~\cite{mcintosh2019android}, vulnerability analysis, and reverse engineering~\cite{pathak2012energy, linares2014mining}. Static and dynamic methods analyze instruction patterns or runtime behavior~\cite{hindle2012green, wan2016detecting}, and large-scale studies have explored code size optimization for native apps~\cite{liu2023linker}. However, these techniques typically focus on code behaviors rather than inferring compilation settings from binary characteristics.

BinEye~\cite{yang2019bineye} reports high accuracy for single-optimization binaries, but does not target mixed-optimization libraries or library-level aggregation at production mobile app scale. Existing methods also rely on source access or controlled settings, which limits third-party app analysis. This distinction matters because app stores expose only packaged binaries, so practical diagnosis must work source-free while handling mixed libraries reused across many production apps. Our work provides binary-only detection with weighted aggregation and ecosystem-scale measurement.

\textbf{Mobile App Performance Analysis.}
Existing mobile app performance analysis primarily focuses on high-level factors such as UI responsiveness, memory leaks, and network latency~\cite{weichbroth2025usability, liao2024automatically}. Tools like Android Profiler concentrate on runtime profiling of Java/Kotlin code and detecting memory bottlenecks~\cite{tang2021systematical, yang2015gator}.

Recent research has explored energy consumption analysis~\cite{flores2024enhancing, hao2013estimating, li2013calculating}, energy-aware design patterns~\cite{cruz2017catalog}, and power modeling tools~\cite{Fieni2024}. Other studies have investigated automated GUI testing~\cite{hu2024enhancing}, energy issue detection~\cite{li2020detecting}, and device-specific behaviors~\cite{dong2024same}. These tools and techniques primarily operate at the application or system level, analyzing runtime metrics and resource usage patterns without examining the underlying compilation configurations that fundamentally determine code efficiency.

Large-scale mobile app studies have also examined app-packaged on-device models~\cite{hu2024first} and cross-device GUI datasets~\cite{hu2023pairwise}, further showing the value of studying deployed mobile apps beyond source-level code. Our focus is complementary: we analyze native-library compiler optimization levels in packaged apps.

Unlike existing work that focuses on application-level optimizations, our study provides the first large-scale empirical investigation of compiler-level inefficiencies in production apps.

\textbf{Compiler Optimization Impact Studies.}
Studies on compiler optimization impact have traditionally focused on controlled benchmark environments and synthetic workloads~\cite{palomba2018beyond, jabbarvand2016energy}. Research has quantified the effects of various optimization techniques on CPU performance metrics and explored trade-offs between optimization levels and compilation time~\cite{manotas2014seeds}. These studies typically measure performance improvements in isolation, using standard benchmark suites or custom test programs compiled with known optimization settings, which limits their applicability to understanding real-world production scenarios.

These studies are typically conducted in controlled settings and do not address prevalence in production apps, third-party root causes, or end-user perception.

\textbf{Remediation Techniques and Complementary Tools.}
Once optimization problems are detected, several techniques can remediate them. Profile-guided optimization (PGO)~\cite{pettis1990profile}, post-link optimization with BOLT~\cite{panchenko2019bolt}, and link-time optimization (LTO)~\cite{lattner2004llvm} improve generated or linked binaries, while recent LLM-based compiler tools further expand automated remediation~\cite{cummins2024large}. These approaches are complementary to \textsc{OptDetect}, which identifies low-optimization libraries and serves as a detection front-end for downstream remediation pipelines.

\section{Conclusion}

This paper presents \textsc{OptDetect}, a source-free framework for detecting compiler optimization problems in mobile applications. Applying it at ecosystem scale, we reveal that 30.5\% of 21,972 native libraries in top-ranked apps are compiled with low optimization levels, affecting 91.7\% of apps. Our automated detection framework achieves 93.0\% accuracy on controlled datasets and 81.9\% on real-world binaries.

Across 12 production apps, fixing detected issues reduces CPU instructions by 10-63\% (median: 20.5\%) for commercial apps and 15-58\% (median: 32\%) for open-source apps. In the 6 commercial apps, complaints and ratings improve in 5 cases, with median changes of 42\% and 0.14 stars. Root cause analysis shows that many issues originate from third-party libraries, and 49.7\% of 1,073 libraries in a major repository exhibit low optimization levels.

Future work includes extending detection to additional architectures and validating with end-to-end runtime energy measurements. In practice, \textsc{OptDetect} can serve as a CI/CD build gate and support pre-publication vetting in SDK repositories.

\begin{acks}
The authors used LLM-based tools for language editing. All technical content and conclusions remain the authors' responsibility.
\end{acks}

\section*{Data Availability Statement}

The \textsc{OptDetect} artifact, including the executable tool, usage instructions, and experimental results, is archived on Zenodo~\cite{optdetect2026artifact}. The source code of \textsc{OptDetect}/ArkAnalyzer-HapRay is publicly available in our \href{https://gitcode.com/SMAT/ArkAnalyzer-HapRay}{GitCode repository}.

\bibliographystyle{ACM-Reference-Format}
\bibliography{reference}

\end{document}